\documentclass[pre,aps,showpacks,twocolumn]{revtex4}
\usepackage{graphicx}
\usepackage{colordvi}
\usepackage{color}
\begin{document}
\title{Normal mode oscillations of a nonlocal composite matter wave soliton}
\author{Mario Salerno$^1$ and Bakhtiyor B. Baizakov$^2$}
\affiliation{$^1$ Dipartimento di Fisica "E.R. Caianiello,"
and INFN Gruppo Collegato di Salerno, \\ Universita di Salerno, Via
Giovanni Paolo II, 84084 Fisciano, Salerno, Italy \\
$^2$ Physical-Technical Institute, Uzbek Academy of Sciences,
100084, Tashkent, Uzbekistan}
\date{\today}

\begin{abstract}
The existence of stable bound states of three solitons in a
Bose-Einstein condensate with nonlocal interactions is demonstrated
by means of variational approach (VA) and numerical simulations. The
potential of interaction between solitons derived from VA is shown
to be of molecular type, i.e. attractive at long distances and
repulsive at short distances. Normal modes of a three-soliton
molecule are investigated by computing small amplitude oscillations
of individual solitons near their equilibrium positions. Symmetric
and asymmetric stretched states of the molecule are prepared and
used as initial conditions in numerical simulations of the nonlocal
Gross-Pitaevskii equation. As opposed to usual triatomic molecules,
we find that the frequency of the asymmetric mode of a three-soliton
molecule is smaller than the one of the symmetric mode. Possible
experimental settings for the observation of these results are
briefly discussed.
\end{abstract}
\pacs{67.85.Hj, 03.75.Kk, 03.75.Lm}
\maketitle

\section{Introduction}

The interaction between solitons and formation of their bound states
has been the subject of long-standing interest in the physics of
nonlinear waves. Some features of soliton interactions, which were
theoretically predicted long ago, are finding confirmation in
present-day experiments~\cite{nguen2014, nguen2017}. The
experimental demonstration of stable two- and three-soliton
complexes, so called {\it soliton molecules}, in dispersion-managed
optical fibers~\cite{stratmann2005,rohrmann2012,rohrmann2013} and
revealing their perspectives for advanced optical telecommunications
\cite{mitschke2016}, has been a notable progress in this direction.
Soliton interactions are important both from the viewpoints of
fundamental physics and practical applications. Motivation for
optical communications has led to the discovery of soliton
interactions in fibers in early years of research on optical
solitons \cite{mitschke1987}. For example, it is known  in soliton
based fiber optic communication
lines~\cite{hasegawa1995,agrawal2002,mitschke2009} that the
interaction of co-propagating solitons can reduce the overall
performance of the system.

Another physical medium, where solitons can exist, is the
Bose-Einstein condensate (BEC) of a diluted atomic gas. Experimental
and theoretical research on solitons in BEC have been reported in
many publications (see review
articles~\cite{abdullaev2005,billam2013,bagnato2015}). Majority of
papers are devoted to properties of single solitons and soliton
trains. Evidence on the interactions between matter-wave solitons
was inferred form the behavior of neighboring solitons, oscillating
in a quasi-1D harmonic trap~\cite{strecker2002,nguen2017}.
Collective dynamics of a chain of solitons, confined by external
potential, in the adiabatic limit has been investigated
in~\cite{gerdjikov2006}. Similar phenomena in two-component BEC were
studied in~\cite{novoa2008}. Regimes to produce bound states of
matter - wave solitons from their collisions were found in
\cite{martin016}. It should be noted, that systematic investigation
of the interaction process of just two or three solitons in BEC
requires precise production and manipulation techniques, which is
being developed nowadays~\cite{marchant2013,medley2014,aycock2017}.

An important fact to be stressed here is that in the mean field
description of BEC, in terms of the Gross- Pitaevskii equation (GPE)
with usual contact atomic interactions, solitons cannot form stable
bound states with finite binding energy. The interaction force
between them depends on the phase difference, and can be either
attractive, or repulsive, and their interaction potential is not of
molecular type. Potential curves for two colliding nonlinear
Schr\"odinger solitons was calculated in \cite{baron2014}. Soliton
complexes in this model, therefore, do not feature a fixed
equilibrium distance, analogous to the bond length of atomic
molecules. In this respect, it is worth mentioning that a breather
consisting of two equal and in-phase solitons, periodically passing
through each-other, as predicted by standard nonlinear Schr\"odinger
equation (NLSE) with focusing cubic nonlinearity, has not been found
in experiments~\cite{mitschke1987}. The reason is that when the two
solitons merge, higher order nonlinear phenomena come into play,
which are not captured by the standard NLSE.

The situation is different in BEC with long range dipole-dipole
atomic interactions. In qualitative terms one can say, that in
dipolar BEC the atoms within one soliton can directly interact with
atoms inside another soliton, so that combined dipolar and usual
phase-dependent interactions of solitons may open the way towards
formation of true matter-wave soliton molecules. The existence of
stable bound states of bright matter-wave solitons in dipolar BEC,
where solitons attract each other at long distances and repel at
short distances, has been theoretically predicted in several papers.
Specifically, soliton bound states in a stack of quasi-1D and
quasi-2D traps were reported in~\cite{lakomy2012}
and~\cite{nath2007}, respectively. In these models individual
solitons, forming the bound state, reside in separate stacks.
Existence of bright solitons and dark-soliton pairs in a dipolar
Tonks - Girardeau gas was investigated in \cite{BAMS2009}. Numerical
analysis of soliton bound states in quasi-2D and 3D dipolar BEC were
also reported in~\cite{lashkin2007,adhikari2011}. Formation of bound
states of solitons and their fusion, resulted from collision of
dipolar solitons have been investigated in \cite{edmonds2017}.  The
vibration spectrum of a two-soliton molecule in dipolar BEC,
confined to a single quasi-1D trap, was studied in
\cite{turmanov2015}, while the potential of interaction, formation
of two-soliton molecules and their binding energy in one dimensional
dipolar BEC were studied by variational approach and numerical
simulations in~\cite{baizakov2015}. Dark solitons in dipolar BEC,
interacting with each other via molecular type potentials and
capable of forming stable bound states, were recently reported
in~\cite{bland2015,pawlowski2015}. Soliton bound states and clusters
in nonlocal optical media are also intensively investigated (for a
recent review see the book~\cite{assanto-book}). Once a soliton
molecule has been created, many interesting phenomena, similar to
those observed in molecular physics, can be modeled with them.

Our main objective in this paper is to study the dynamics and normal
mode oscillations of three-soliton molecules, which can exist
in nonlocal media. To this end we develop a variational approach
(VA)~\cite{anderson1983, malomed2002} to find the stationary shape
of a three-soliton molecule, reveal the character of the interaction
potential and estimate the frequency of small amplitude oscillations
of solitons near their equilibrium positions. VA stationary profiles
of three-soliton molecules are also found in very good agreement
with the numerical ones obtained from a self-consistent (SC)
procedure \cite{salerno2005} applied to the GPE. To explore the
molecular dynamics we prepare symmetric and asymmetric stretched
states of the molecule by imposing constant and non-uniform chirping
of the ground state wave function and used them as initial
conditions for the GPE, and recording the positions of each soliton
during their time evolution, constitutes the basis of our numerical
experiments.

As a result we show that when considered in proper coordinates the
GPE dynamics simplifies, displaying harmonic oscillations which
resemble the ones of normal modes of usual triatomic linear
molecules. In contrast to what observed in molecular physics,
however, we find that the oscillation frequency of the motion
induced from a symmetric stretching  is always larger than the one
induced from an asymmetric stretching of the three-soliton molecule.
We find that the VA predictions for stationary three-soliton
molecules and for the symmetric oscillations of the molecule are in
excellent agreement with numerical GPE integrations. The VA,
however,  does not allow to make predictions for asymmetric
oscillations due to the difficulty of finding suitable trial
functions for this case. Normal mode oscillations have also been
investigated for topological soliton bound states of the sine-Gordon
equation~\cite{SS88} and for the displaced dynamics of binary BEC
mixtures~\cite{GS12}.

The paper is organized as follows. In the next Sec. II we develop
the VA using the Gauss-Hermite trial function for a three-soliton
molecule and check its validity by comparing its predictions with
the results of numerical solution of the governing nonlocal GPE.
In Sec. III we consider initially deformed states suitable to
excite internal mode oscillations of the molecule and use them as
initial conditions for numerical integrations of the GPE. Results
are then compared with predictions of the VA analysis. In Sec. IV
we briefly summarize our findings and discuss the generality of
our results with respect to other types of nonlocal interactions.
Possible experimental settings and  areas of research where the
obtained results might be useful, are also briefly discussed.

\section{Model Equations and variational analysis}

The governing equation of our model is a 1D nonlocal
Gross-Pitaevskii equation, represented in normalized units as
follows
\begin{equation} \label{gpe}
i\frac{\partial \psi}{\partial t} + \frac{1}{2} \frac{\partial^2
\psi}{\partial x^2}+ q |\psi|^2 \psi + g \psi \int
\limits_{-\infty}^{+\infty} R(|x-\xi|) \, |\psi(\xi,t)|^2 d \xi=0,
\end{equation}
where $\psi(x,t)$ is the mean field wave function of the condensate,
$q$ and $g$ are coefficients of nonlinearity, responsible for the
local contact and long-range nonlocal atomic interactions,
respectively. The wave function is normalized to the number of atoms
in the condensate $N=\int_{-\infty}^{+\infty} |\psi(x)|^2 dx$, which
is a conserved quantity of Eq.~(\ref{gpe}). Since the nonlocal
interaction is essential for the existence of soliton molecules and
molecular dynamics, we shall concentrate mainly on the case $q=0$,
and discuss at the end that results may be preserved also in the
presence of contact interactions. We also remark, that in
experiments it is possible to de-tune the cubic nonlinearity to zero
by means of Feschbach resonances~\cite{chen2006}.

The response function $R(x)$ in Eq. (\ref{gpe}) characterizes the
degree of nonlocality of the medium, which shows how strongly the
properties at a given location depend on the properties of its
neighborhood. For analytical convenience we consider a Gaussian
function normalized to one
\begin{equation}\label{kernel0}
R(x) = \frac{1}{\sqrt{2\pi} w} \exp \left(-\frac{x^2}{2w^2} \right),
\end{equation}
and show in the last section that similar results can be obtained
also for long-ranged response functions with algebraic, instead of
exponential, decay at large distances.
\begin{figure}[htb]
\centerline{\includegraphics[width=4.4cm,height=4.4cm,clip]{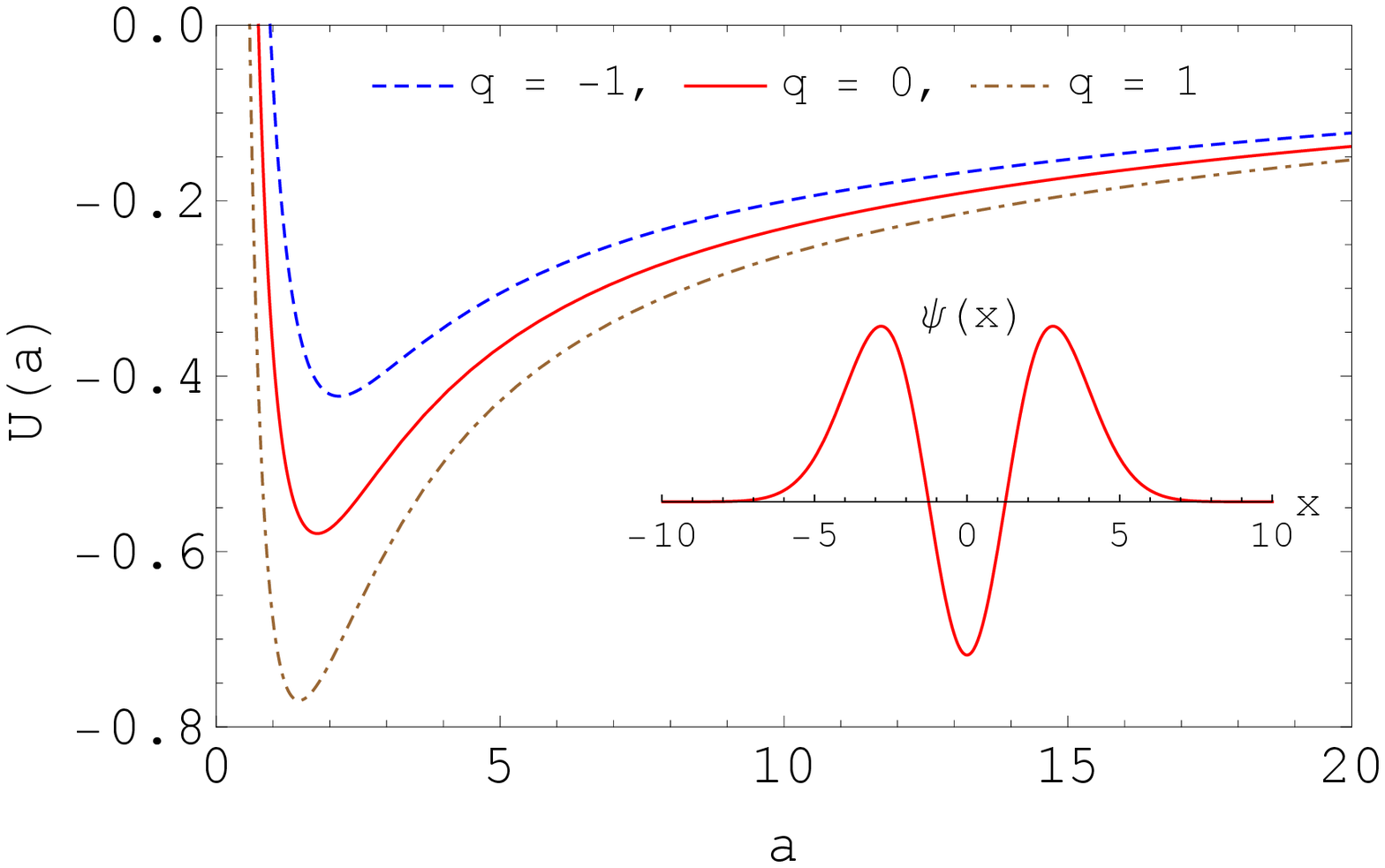}
            \includegraphics[width=4.4cm,height=4.4cm,clip]{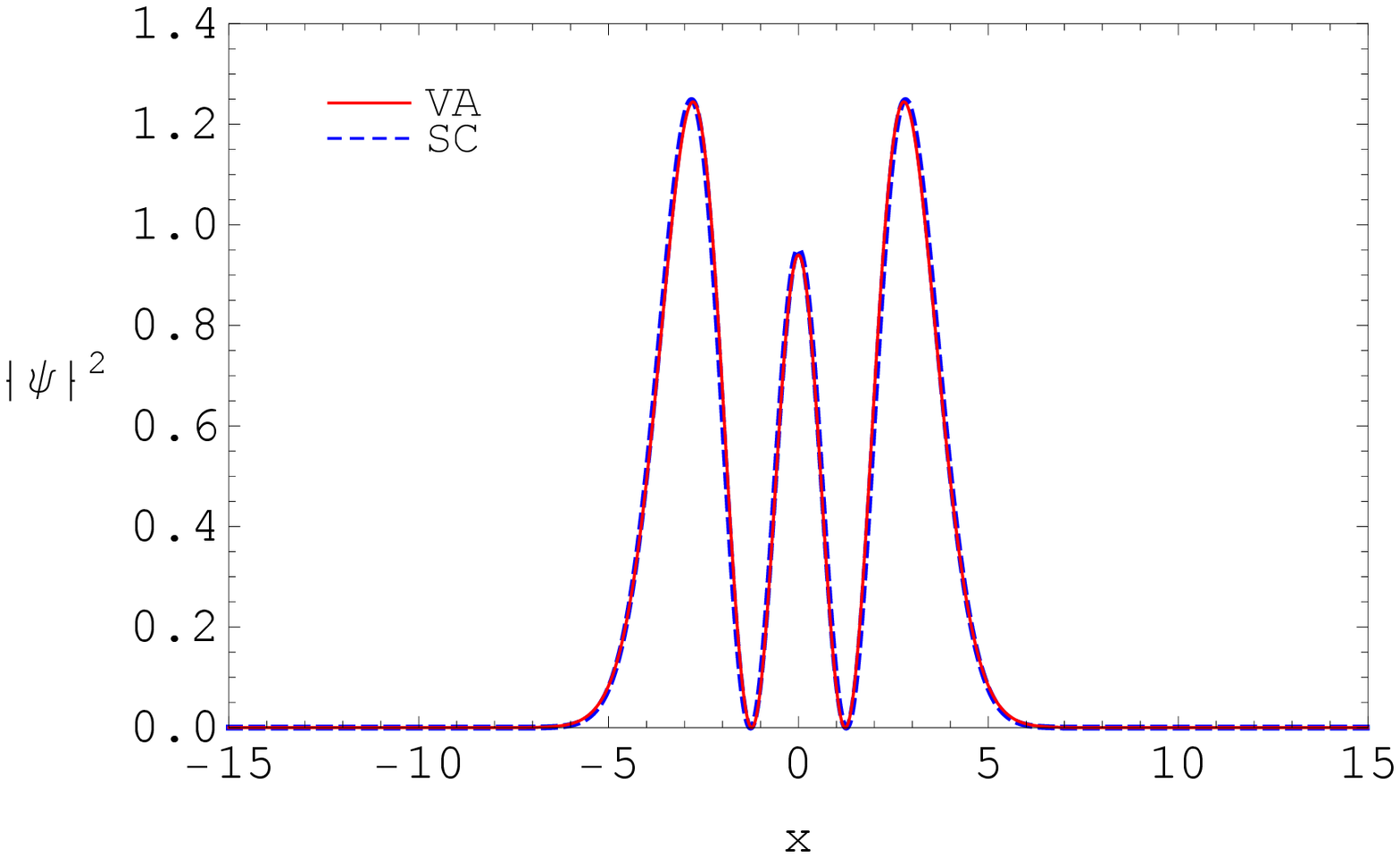}}
\caption{Left panel: The potential $U(a)$ for $g=10$ and $q=0, \,
\pm \, 1$. The inset shows the wave function of a three-soliton
molecule for $q=0$. Right panel: The modulo square of the wave
function, according to VA for parameter values $N=6$, $w=5$,
$q=0$, $g=10$, $A=0.975$, $a=1.781$. Dashed line represents the
stationary wave profile, constructed using the self-consistent
procedure~\cite{salerno2005}. The two curves nearly coincide,
which evidences that Eq. (\ref{ansatz}) represents a good trial
function. } \label{fig1}
\end{figure}
The parameter $w$ in Eq. (\ref{kernel0}) designates the
strength of nonlocality. At $w \rightarrow 0$ the response
function resembles the Dirac delta function. In this case the
medium is called weakly nonlocal. In the opposite case of large
$w$, compared to the waist of the excitation, the medium is called
highly nonlocal. The response function for a dipolar BEC,
confined to quasi-1D trap, was derived in~\cite{sinha2007}.

For three-soliton bound states we can employ the variational
approach similar to that developed in~\cite{baizakov2015,
turmanov2015}. As a suitable trial function we use the second
Gauss-Hermite function
\begin{equation} \label{ansatz}
\psi(x,t)=A\left(2\frac{x^2}{a^2}-1\right) \exp \left[
-\frac{x^2}{2a^2} + i b x^2 + i\phi \right],
\end{equation}
where the variational parameters $A(t), a(t), b(t), \phi(t)$ have
the meaning of amplitude, width, chirp and phase, respectively.
It should be noted, that this waveform can be modelled by
three Gaussian functions, arranged in anti-phase configuration.
When the phase difference between adjacent solitons differs from
$\phi= \pi$, stable bound state of three solitons does not emerge,
as we have found from numerical simulations.

The norm of the trial function, which is proportional to reduced
number of atoms, is $N~=~2 A^2 a \sqrt{\pi}$. To develop the VA we
note that Eq.~(\ref{gpe}) can be obtained from the Lagrangian
density:
\begin{eqnarray}\label{lagrangian}
{\cal L} &=& \frac{i}{2}(\psi \psi^{\ast}_t - \psi^{\ast}\psi_t) +
\frac{1}{2} |\psi_x|^2 - \frac{1}{2} q |\psi|^4  \nonumber\\ && - \frac{1}{2} g
|\psi(x,t)|^2 \int \limits_{-\infty}^{\infty} R(x-\xi)
|\psi(\xi,t)|^2 d \xi.
\end{eqnarray}
Using the response function (\ref{kernel0}) and the ansatz in
Eq.~(\ref{ansatz}), we evaluate the Lagrangian density
(\ref{lagrangian}).
Subsequent integration over the space variable $L= \int {\cal L} dx$
yields the averaged Lagrangian
\begin{equation}
\frac{L}{N} = \frac{5}{2} a^2 b_t + \phi_t + \frac{5}{4a^2} + 5\,a^2
b^2 - \frac{41 q N}{128 \sqrt{2\pi} a} - \frac{g N}{2 \sqrt{2\pi}}
F(a,w),
\label{lagrangian2}
\end{equation}
where
\begin{equation}\label{Faw}
F(a,w)=\frac{w^8+2w^6a^2+\frac{15}{4}w^4a^4+\frac{7}{4}
w^2a^6+\frac{41}{64}a^8}{(w^2+a^2)^{\frac{9}{2}}}.
\end{equation}
\begin{figure}[htb]
\centerline{\includegraphics[width=4.7cm,height=4.7cm,clip]{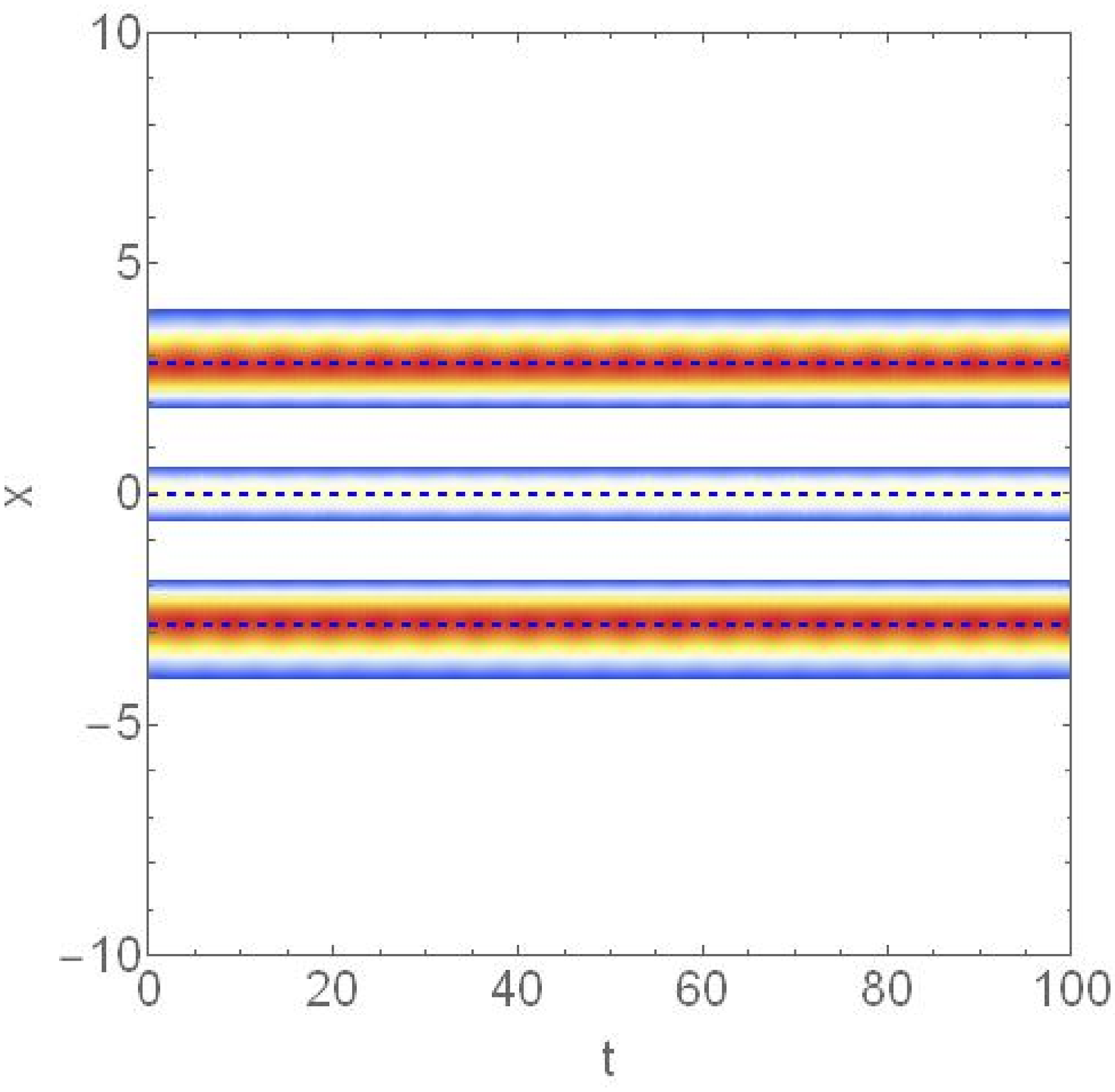}
            \includegraphics[width=4.7cm,height=4.7cm,clip]{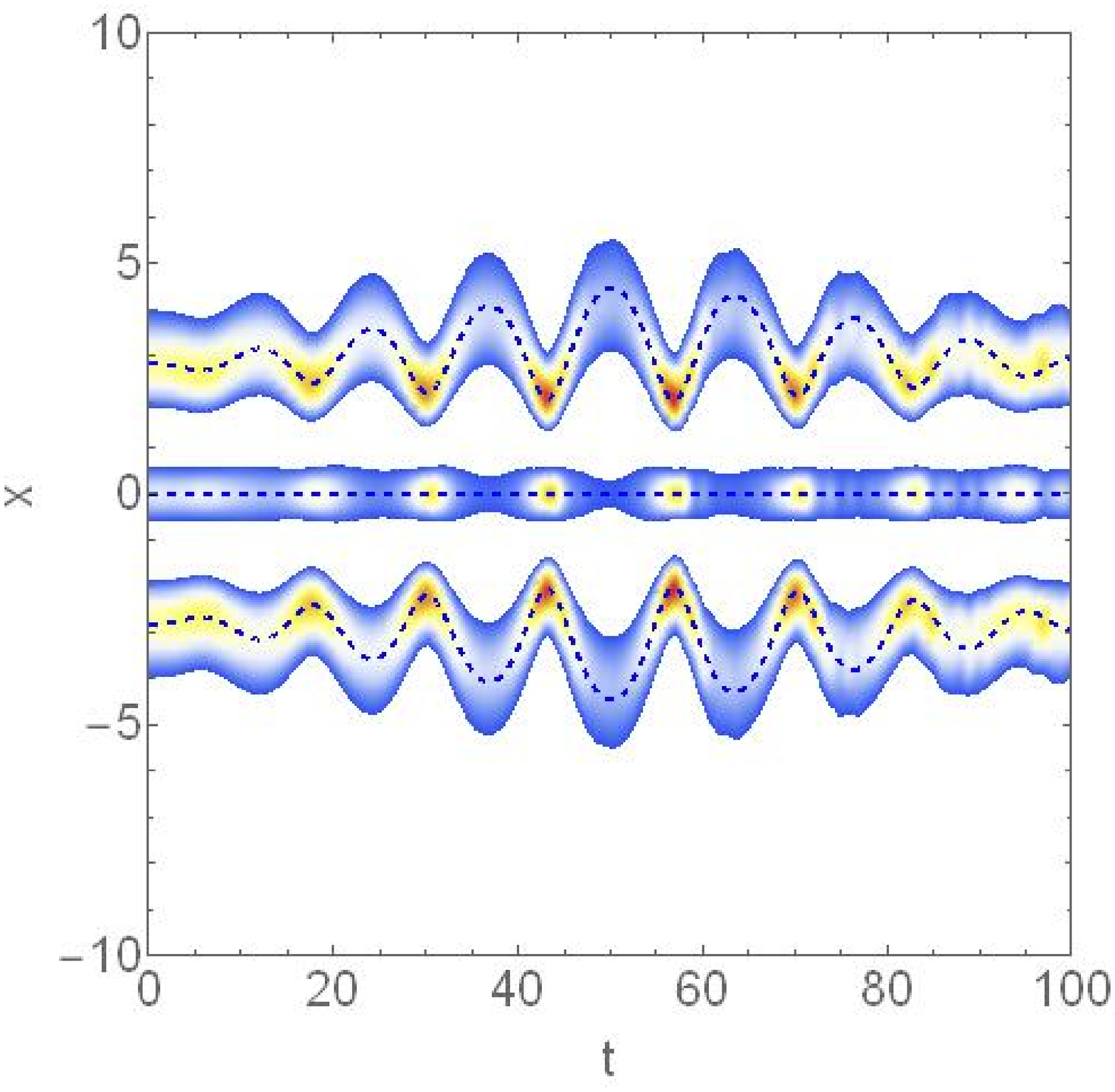}}
\caption{Left panel: Stable propagation of the three-soliton
molecule with parameters predicted by VA. The density plot is
obtained by numerical solution of the GPE (\ref{gpe}). Dashed
lines correspond to positions of maxima of the central and lateral
solitons $x_m=\pm \, \sqrt{5/2}\,a(t)$, where the time dependent
parameter $a(t)$ is evaluated from Eq.~(\ref{att}). Right panel:
Periodic variation of the strength of dipolar interactions
$g(t)~=~g_0(1+\epsilon \sin(\omega_0 t))$ at resonant frequency
$\omega_0=0.529$ gives rise to vibration of lateral solitons with
growing amplitude, while the central soliton remains at origin due
to the symmetry. Parameter values: $g_0=10$, $\epsilon=0.1$. Other
parameters are similar to Fig.~\ref{fig1}. } \label{fig2}
\end{figure}
From the Euler-Lagrange equations \\ $d/dt(\partial L/\partial
\nu_t)-
\partial L/\partial \nu = 0$ for the variational parameters
$\nu~\rightarrow~a,~b,~\phi,$ we obtain the following equation for $a(t)$
\begin{equation}\label{att}
a_{tt}=\frac{1}{a^3}-\frac{41\,q\,N}{320\sqrt{2\pi} a^2} +
\frac{g\, N}{5\sqrt{2\pi}}\,\frac{\partial F(a,w)}{\partial a}.
\end{equation}

This equation has formal analogy with the equation of motion for a
unit mass particle performing oscillations in the anharmonic
potential $U(a)$:
\begin{equation}\label{pot}
U(a)=\frac{1}{2a^2}-\frac{41\,q\,N}{320 \sqrt{2\pi}\,a}
-\frac{g\,N}{5\sqrt{2\pi}}\, F(a,w),
\end{equation}
depicted in Fig.~\ref{fig1}. The minimum of the potential
(\ref{pot}) at $a = a_0$ corresponds to stationary width of the
molecule. The frequency of small amplitude oscillations of the
molecule can be estimated from $\omega_0^2 =
\partial^2 U/\partial a^2 |_{a\rightarrow a_0}$. It should be
pointed out that the interaction potential between solitons, given
by Eq. (\ref{pot}), is of a molecular type, i.e. solitons attract
each other at long distance ($\partial U/\partial a |_{a > a_0} >
0$), and repel at short distance ($\partial U/\partial a |_{a < a_0}
< 0$), so that if $a>a_0$ the distance between solitons tends to
shrink, and for $a~<~a_0$ tends to expand. At the equilibrium
distance attractive and repulsive forces balance each other, and the
solitons remain motionless. In the right panel of Fig. \ref{fig1} we
show the stationary wave profile, found from the fixed point of
Eq.~(\ref{att}) and compare it with the exact wave profile
numerically obtained from a self-consistent (SC)
procedure~\cite{salerno2005} applied to GPE (\ref{gpe}). The
excellent agreement confirms the validity of the trial function in
Eq. (\ref{ansatz}) for  our analytical calculations.

In analogy with the bond length of ordinary molecules composed of
neutral atoms, the distance between maxima of two lateral solitons
$\Delta =2 x_m = \sqrt{10} \, a$, can be a characteristic
parameter of the soliton molecule.
\begin{figure}[htb]
\centerline{\includegraphics[width=6cm,height=6cm,clip]{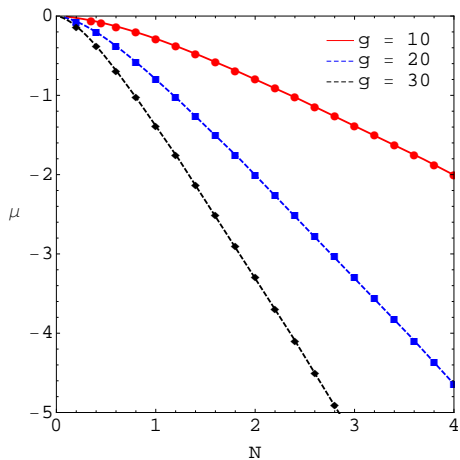}}
\caption{The chemical potential as a function of the norm for
three values of the nonlocal coefficient and $q=0$. The curves are
drawn according to VA Eqs. (\ref{mu}) -- (\ref{norm}), while the
symbols represent the data, obtained by SC
procedure~\cite{salerno2005}. It is evident, that stronger
nonlocal interaction leads to more stable soliton molecules.}
\label{fig3}
\end{figure}

To check the accuracy of the VA we have periodically modulated in
time the strength of the nonlocal nonlinearity, $g(t)$, and compared
the results of Eq.~(\ref{att}) with numerical solution of the
GPE~(\ref{gpe}). In experiments with dipolar BEC, such a {\it
dipolar nonlinearity management} can be implemented by means of
rotating magnetic fields~\cite{giovanazzi2002, tang2018}.
Alternatively, this can be achieved by slowly varying the
polarization angle $\theta$, since $g \sim (1-3 \cos^2\theta)$,
where $\theta$ is the angle between the long axis of the quasi-1D
trap and the dipoles. Fig.~\ref{fig2} illustrates the dynamics of
the three-soliton molecule under varying strength of nonlocal
interaction. As it can be seen from this figure, the VA provides an
accurate description of the dynamics. The stability of localized
solutions for nonlinear wave equations can be examined by means of
the Vakhitov-Kolokolov (VK) criterion~\cite{vakhitov1973}. Following
the usual procedure~\cite{malomed2002} we look for stationary
solutions of the GPE as $\psi(x,t) = \varphi (x) \exp(- i\mu t)$,
where $\mu$ denotes the chemical potential. The time-independent GPE
takes the form
\begin{equation}\label{stat_gpe}
\mu \varphi + \frac{1}{2} \varphi_{xx} + q \varphi^3 + g \varphi
\int \limits_{-\infty}^{\infty} R(|x-z|) \varphi^2(z) dz = 0,
\end{equation}
and the corresponding Lagrangian density is
\begin{displaymath}
{\cal L}=\frac{1}{4}\left[\varphi_x^2- 2\mu \varphi^2 - q \varphi^4
- g \varphi^2 \int \limits_{-\infty}^{\infty} R(|x-z|) \varphi^2(z)
dz\right].
\end{displaymath}
Performing further standard VA procedures with the ansatz
\begin{equation}
\varphi(x) = A \left(2\,\frac{x^2}{a^2}-1\right)
\exp\left(-\frac{x^2}{2 a^2}\right),
\label{VA-stationary}
\end{equation}
and using the response function (\ref{kernel0}),  we get the following
expressions for the chemical potential and norm:
\begin{eqnarray}
\mu &=& -\frac{q\,N}{a}\,\frac{123}{256 \sqrt{2 \pi}} -
\frac{g\,N\,}{\sqrt{2\pi}}
\left(F + \frac{a}{4} \frac{\partial F}{\partial a} \right), \label{mu} \\
N &=& \frac{320 \sqrt{2\pi}}{a \left(41 q - 64 \,g\,a^2
\frac{\partial F}{\partial a}\right)}, \label{norm}
\end{eqnarray}
with the function $F(a,w)$ given by Eq.~(\ref{Faw}).
\begin{figure}[htb]
\centerline{\includegraphics[width=4.7cm,height=4.7cm,clip]{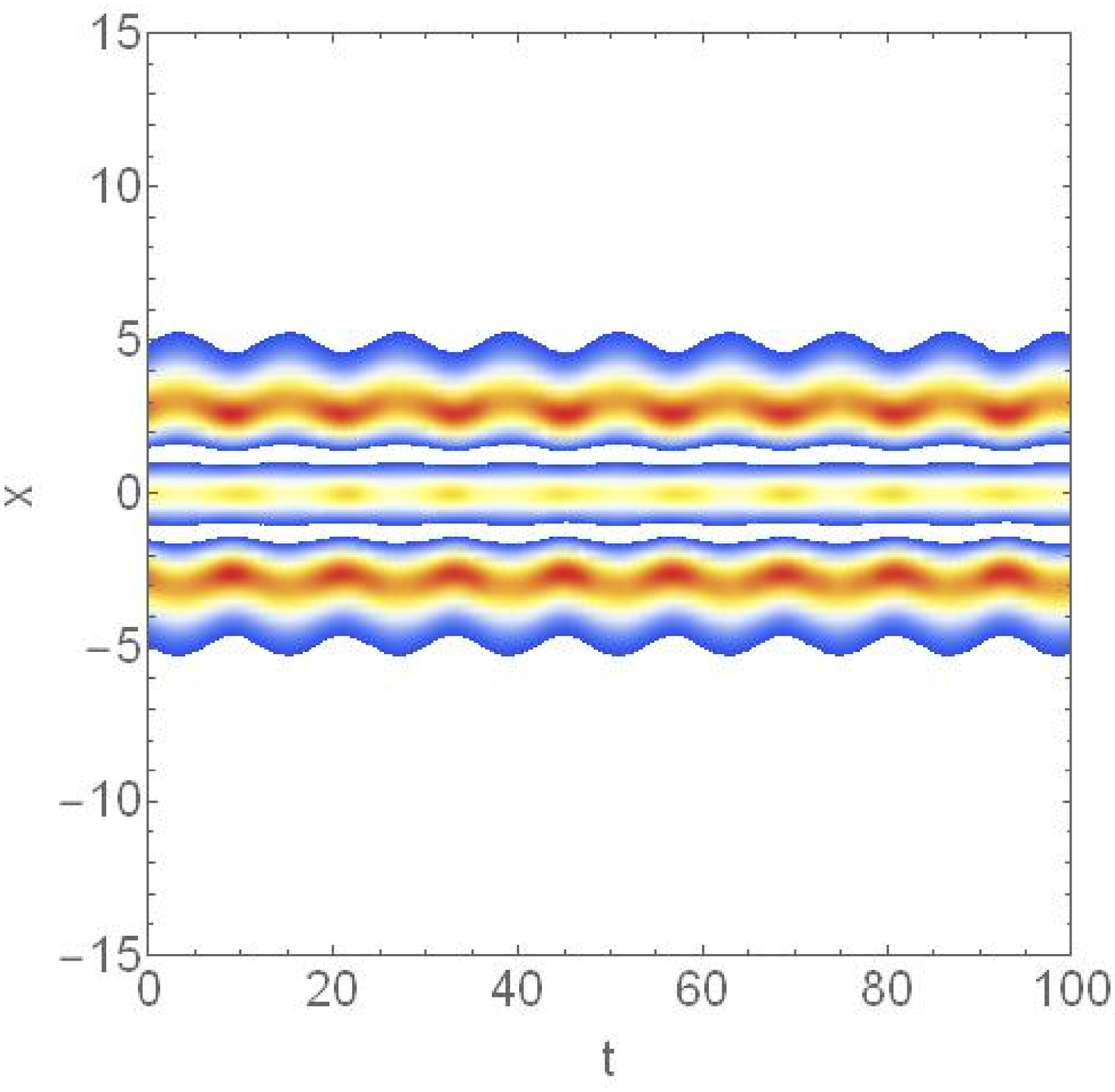}
            \includegraphics[width=4.7cm,height=4.7cm,clip]{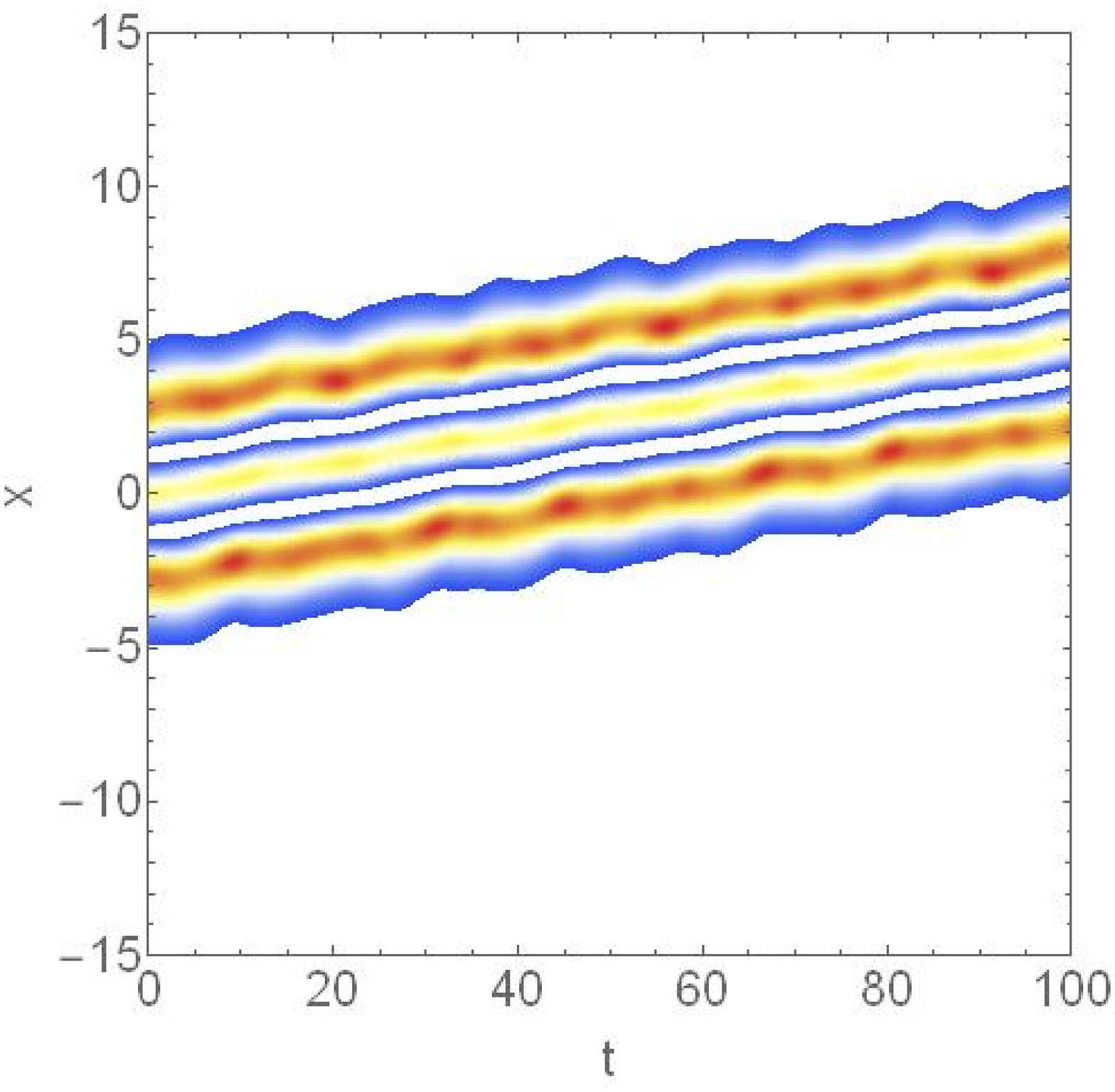}}
\caption{Excitation of the symmetric (left) and both the symmetric
and asymmetric (right) modes of the three-soliton molecule. The
symmetric mode has been excited via constant chirping $\psi_{gs}e^{i
b x^2}$ with $b=0.02$. For the asymmetric mode $b=0$ for $x \leq 0$
and $b=0.02$ for $x>0$.}
\label{fig4}
\end{figure}
From the parametric plot $\mu(a)$ versus $N(a)$ depicted  in
Fig.~\ref{fig3}, one can see that the condition $\frac{d\mu}{d N}
< 0$ is always satisfied, which suggests, according to VK
criterion, the stability of the three-soliton molecule for
different values of the coefficient $g$. As expected, the stronger
attraction between solitons leads to more stability of the
molecule.

\section{Numerical Results}

\begin{figure}[htb]
\centerline{
\includegraphics[width=4.5cm,height=4.5cm,clip]{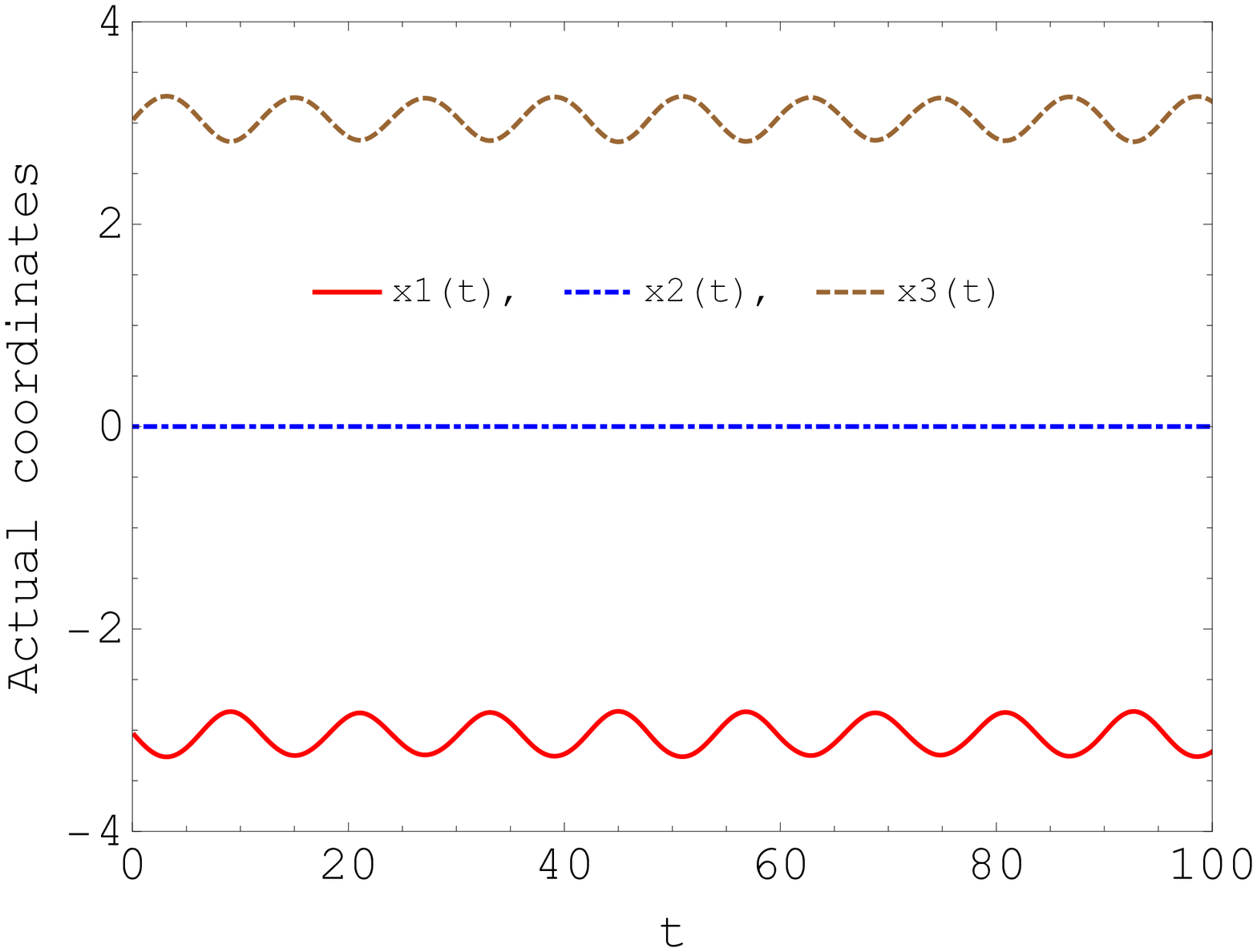}
\includegraphics[width=4.5cm,height=4.5cm,clip]{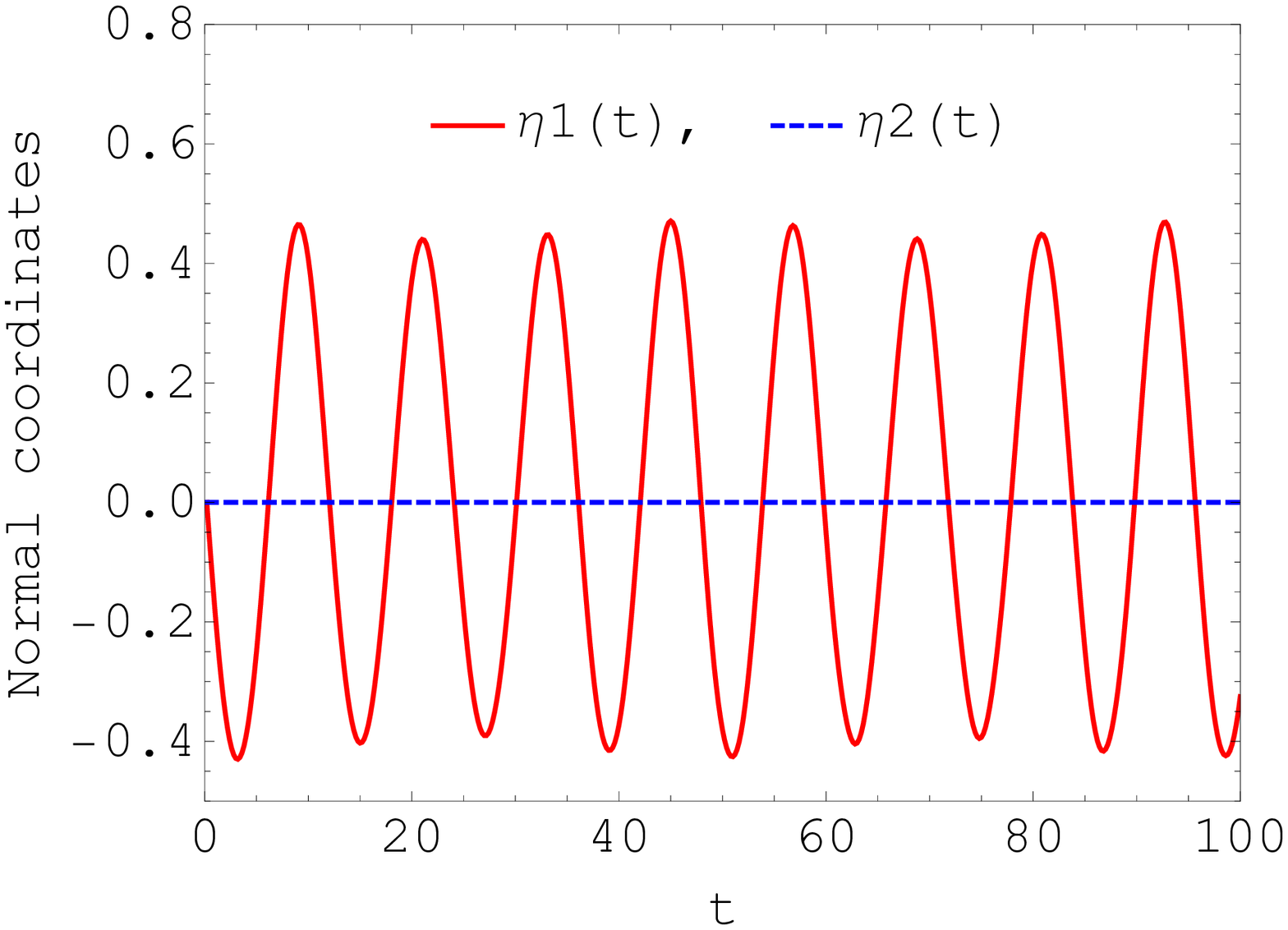}}
\vskip .25cm \centerline{
\includegraphics[width=4.5cm,height=4.5cm,clip]{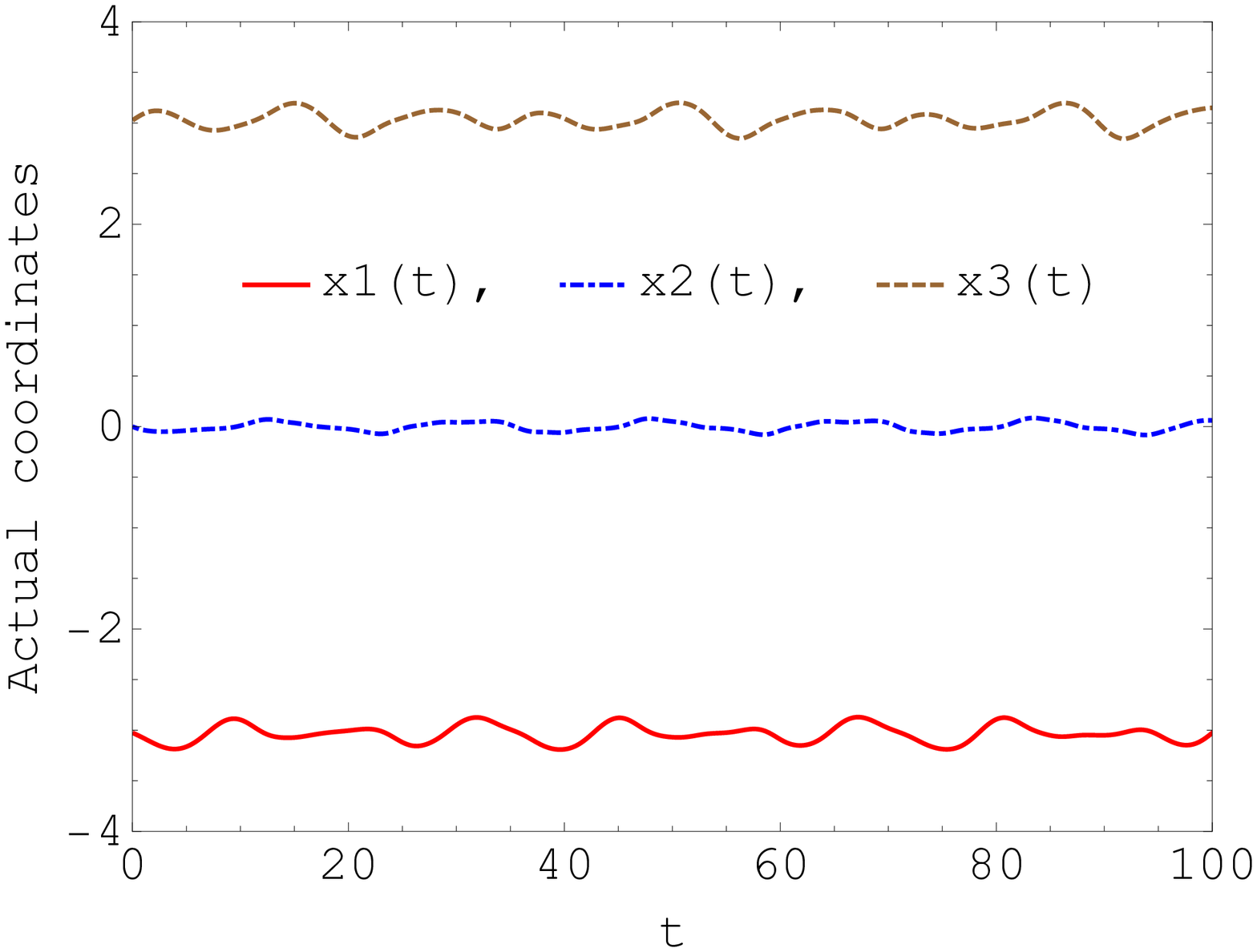}
\includegraphics[width=4.5cm,height=4.5cm,clip]{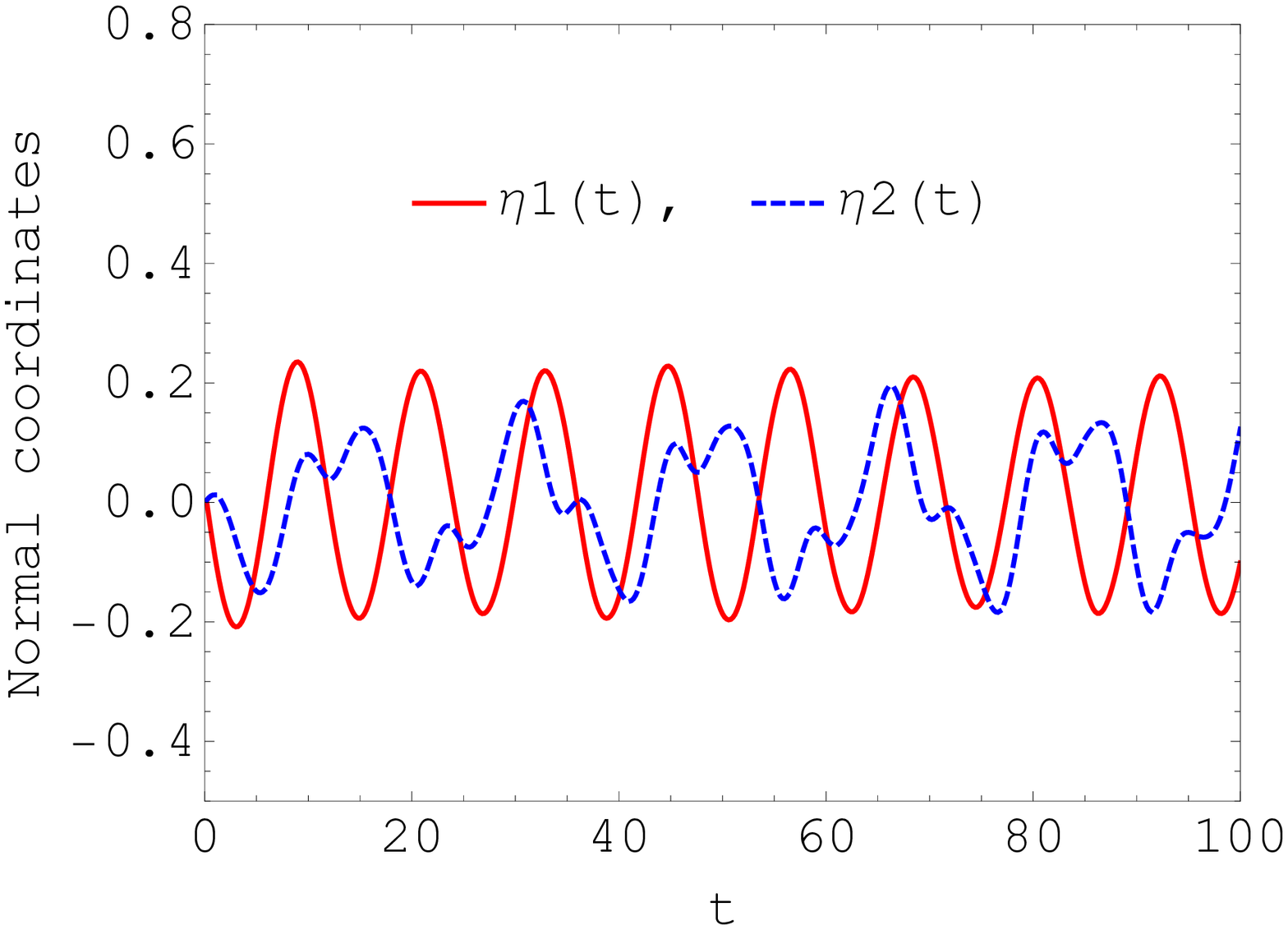}}
\caption{Center of mass positions of individual solitons of a
three-soliton molecule, represented using actual (left panels) and
normal mode coordinates (right panels), according to numerical
solution of Eq. (\ref{gpe}), in the reference frame, attached to the
moving molecule. Small deviations from pure sinusoidal character in
the asymmetric mode (blue dashed line) is due to the matter exchange
between solitons. The normal mode frequencies found from GPE
simulations are equal to $\omega_s= 0.53$, $\omega_a=0.35$ for
parameter values fixed as in Fig.~\ref{fig1}. }
\label{fig5}
\end{figure}

To explore the molecular three-soliton dynamics we need to prepare
initial symmetric and asymmetric stretched states of the molecule,
and use them as initial conditions for numerical simulations of the
GPE. However, stretching and releasing the molecule in such a way,
that each soliton oscillates near its equilibrium position, while
the center of mass of the molecule remains at rest (as usually
presumed by the normal modes theory), is quite challenging problem.
That is why we employ another approach to excite symmetric and
asymmetric modes of the molecule, initially prepared in its ground
state $\psi_{gs}$. In particular, to excite only the symmetric mode
of the molecule, when the flanking solitons oscillate in anti-phase,
while the central soliton does not move, we impose constant chirping
$\psi_{gs}e^{i b x^2}$ with a small chirp parameter $b \ll 1, x \in
[-\infty, \infty]$. To excite both the symmetric and asymmetric
modes, we impose inhomogeneous chirping $b = 0$ for $x<0$, and $b
\neq 0$ for $x>0$. It should be stressed, that inhomogeneous
chirping induces vibration of all solitons, as well as motion of the
entire molecule, as shown in the right panel of Fig. \ref{fig4}.
Below we employ the reference frame, attached to the moving
molecule. In Fig.~\ref{fig5} we show the time evolution of the
center of mass positions of individual solitons of a three-soliton
molecule, excited as described above.

Since the resulting dynamics of the molecule is a superposition of
different modes, its periodic character is not readily recognized in
the actual center of mass coordinates $x_i$ (see lower left panel of
Fig.~\ref{fig5}). The periodic motions, however, become evident if
one introduces the coordinates $\eta_i$ defined as
\begin{equation}
\label{normalcoord} \eta_1(t)=x_1(t)-x_3(t), \qquad
\eta_2(t)=x_1(t)+x_3(t),
\end{equation}
where $x_i(t)$ denote the displacement of the solitons with respect
to their equilibrium positions. Note that apart from constant
factors, these coordinates are just the same as the normal mode
coordinates of usual linear triatomic molecules. Obviously, the
model is valid for small amplitude oscillations of solitons when
anharmonic effects are negligible.

In normal mode coordinates (\ref{normalcoord}) the dynamics indeed
looks periodic, and the frequency of the symmetric and asymmetric
modes are easily identified (see right panels of Fig.~\ref{fig5}).
In this respect, the soliton molecule behaves similar to usual
triatomic molecule. However, there is also a significant difference
between these two systems. It concerns the flow of matter between
solitons during the time evolution, which is considered below.

The mass of each soliton ($m_i, i=1,2,3$) is proportional to its
norm
\begin{displaymath}
m_1(t) = \int \limits_{-\infty}^{z_1(t)} n dx, \ m_2(t) = \int
\limits_{z_1(t)}^{z_2(t)} n dx, \ m_3(t) = \int
\limits_{z_2(t)}^{\infty} n dx,
\end{displaymath}
where $n=|\psi (x,t)|^2$ is the density of the condensate according
to GPE (\ref{gpe}), $z_1(t)$, $z_2(t)$ are the left and right
borders of the middle soliton (where the field amplitude vanishes
$\psi (x,t) \rightarrow 0$).

\begin{figure}[htb]
\centerline{
\includegraphics[width=4.6cm,height=4.cm,clip]{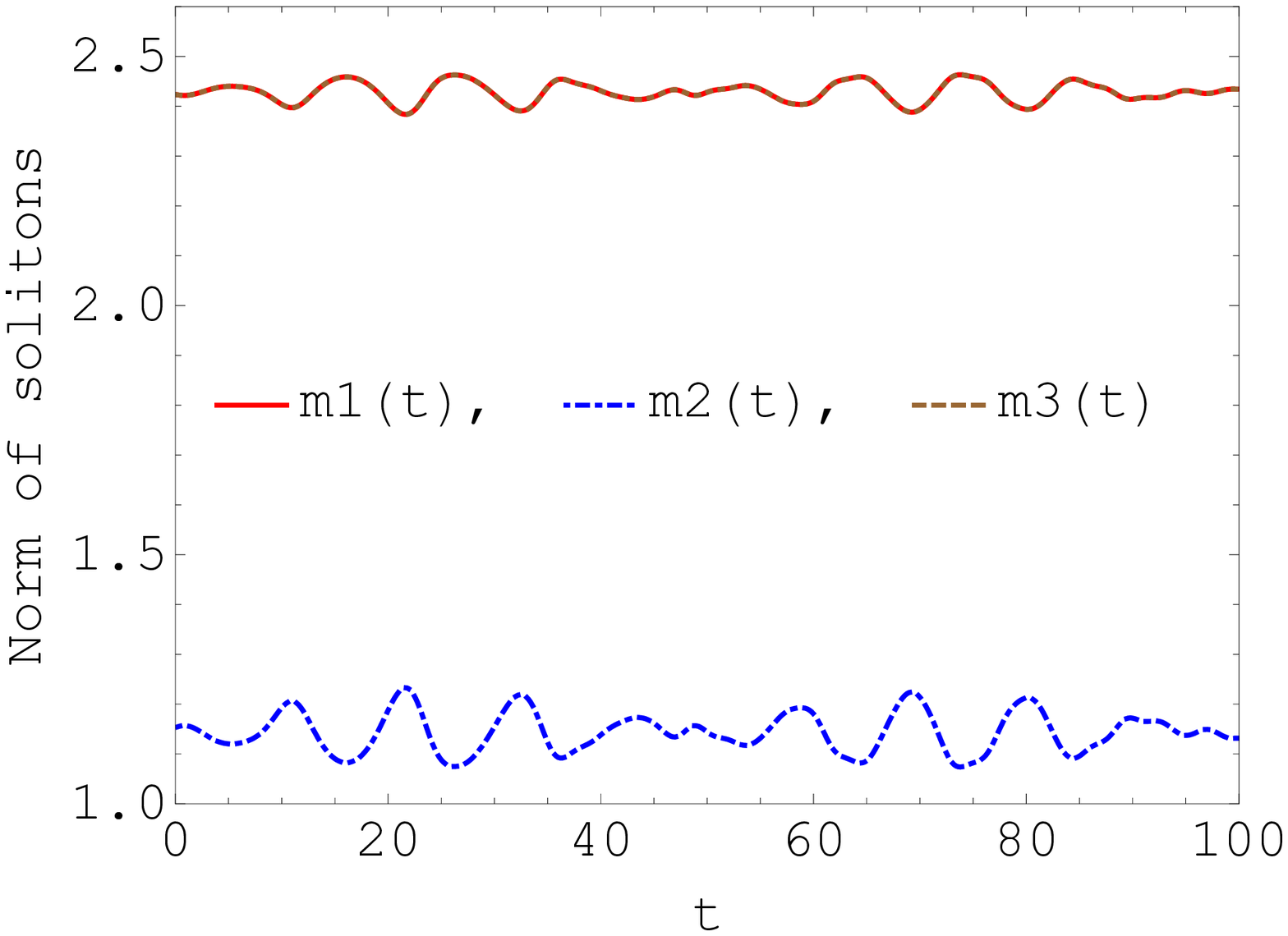}
\includegraphics[width=4.6cm,height=4.cm,clip]{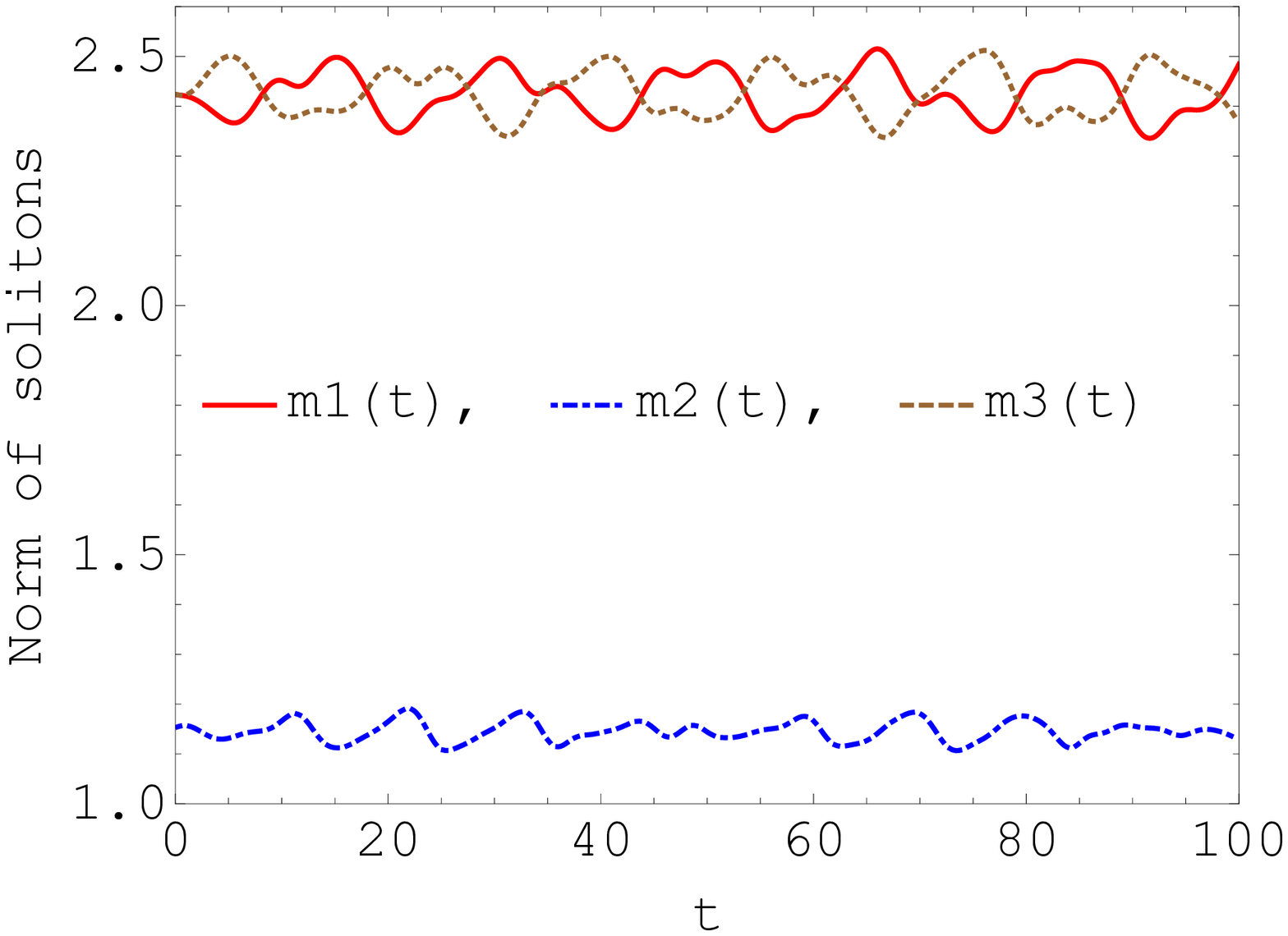}}
\caption{Excitation of the molecule's vibrational modes leads to
small exchange of matter between solitons. The total mass is always
conserved $m_1(t)+m_2(t)+m_3(t)=\mbox{const}$. When the symmetric
mode has been excited (left), flanking solitons exchange equal
amount of matter with the central soliton. In this case the curves
for $m_1(t)$ and $m_3(t)$ coincide. For the asymmetric mode (right),
there is a dynamic imbalance between masses of flanking solitons
$m_1(t)$ and $m_3(t)$.}
\label{fig6}
\end{figure}

Time dependence of these quantities
implies, that each soliton of the molecule periodically
expand/shrink and move. Evaluation of masses of solitons according
to above formulas shows, that there is small exchange of matter
between solitons, when the vibrations of the molecule has been
excited, as illustrated in Fig. \ref{fig6}.
The strength of interaction between solitons, and therefore
vibration frequency of soliton bound states, depends on the number
of atoms (expressed via norm).

The frequency of symmetric oscillations of the molecule can be
predicted by VA through the second derivative of the potential in
Eq.~(\ref{pot})
\begin{equation}\label{VAoms}
\omega_s =\sqrt{\partial^2 U/\partial a^2|_{a \rightarrow a_0}}.
\end{equation}

 We find that this expression leads to results that are in very
good agreement with GPE numerical calculations, as shown below.
For the frequency of asymmetric mode $\omega_a$, however,
analytic estimate is not available.

\begin{figure}[htb]
\centerline{
\includegraphics[width=7cm,height=6cm,clip]{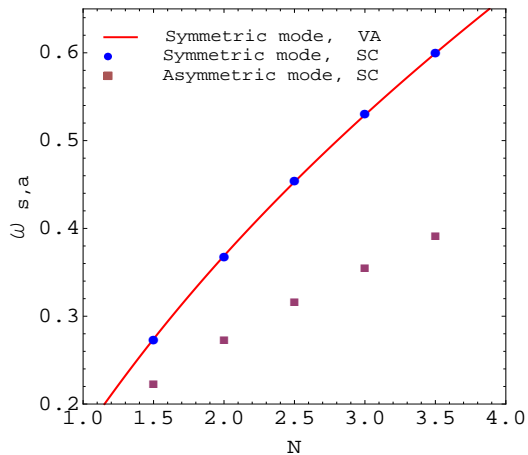}}
\caption{Normal mode frequencies of the three-soliton molecule,
obtained by numerical solution of the GPE (\ref{gpe}) with initial
waveforms, corresponding to different norms of the molecule.
Validity of the relation (\ref{ineq}) is confirmed for all
selected parameters. The symmetric mode frequency is predicted by
the VA expression in Eq. (\ref{VAoms}) (red solid line), while for
the asymmetric mode frequency an analytic estimate is not
available. } \label{fig7}
\end{figure}

Quite interestingly, we find that the frequency of the asymmetric
mode is always smaller than the one of the symmetric mode
\begin{equation}\label{ineq}
\omega_s > \omega_a. \label{omgs}
\end{equation}
It is well known, that in usual triatomic molecules the opposite
relation holds. In Fig. \ref{fig7} the symmetric and asymmetric mode
frequencies are plotted as a function of the norm. From this figure
it is also evident that the numerical results for the symmetric
frequency are in excellent agreement with the ones derived from the
VA expression in Eq. (\ref{VAoms}). Similar behaviors were found for
generic parameter values and for other initial conditions. It is not
simple, however, to account for the asymmetric oscillation frequency
of the molecule by means of the VA. In this respect notice that the
ansatz in Eq. (\ref{ansatz}) does not allow any asymmetric dynamics.

\section{Discussion and Conclusions}

Before closing this paper we feel compelled to discuss in more
details the feasibility of the above results for dipolar BEC, and
possible experimental settings to verify the proposed model. In this
regard, the following remarks are in order.

\subsection{About the response functions}

\begin{figure}[htb]
\centerline{\includegraphics[width=4.7cm,height=4.7cm,clip]{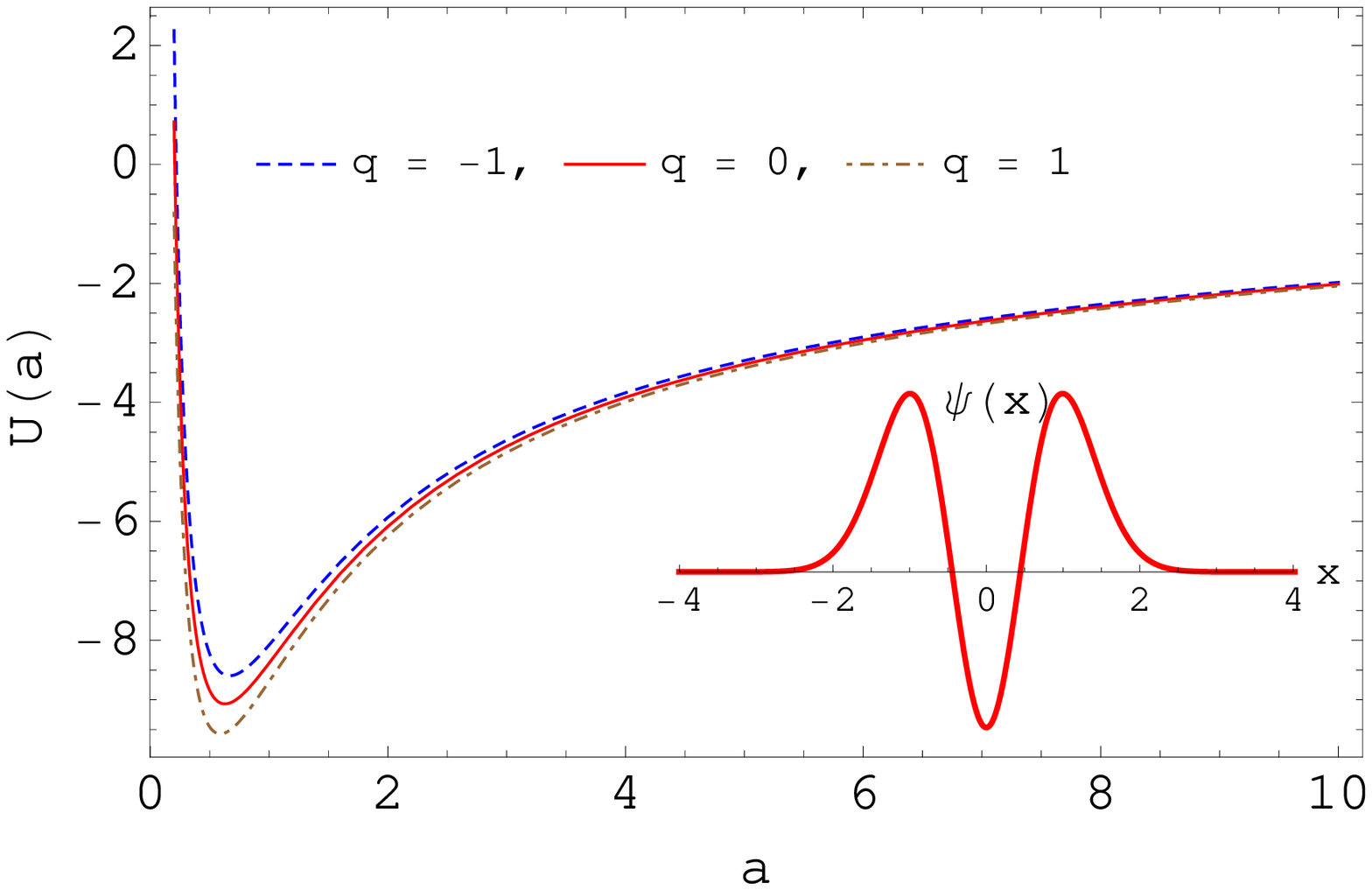}
            \includegraphics[width=4.7cm,height=4.7cm,clip]{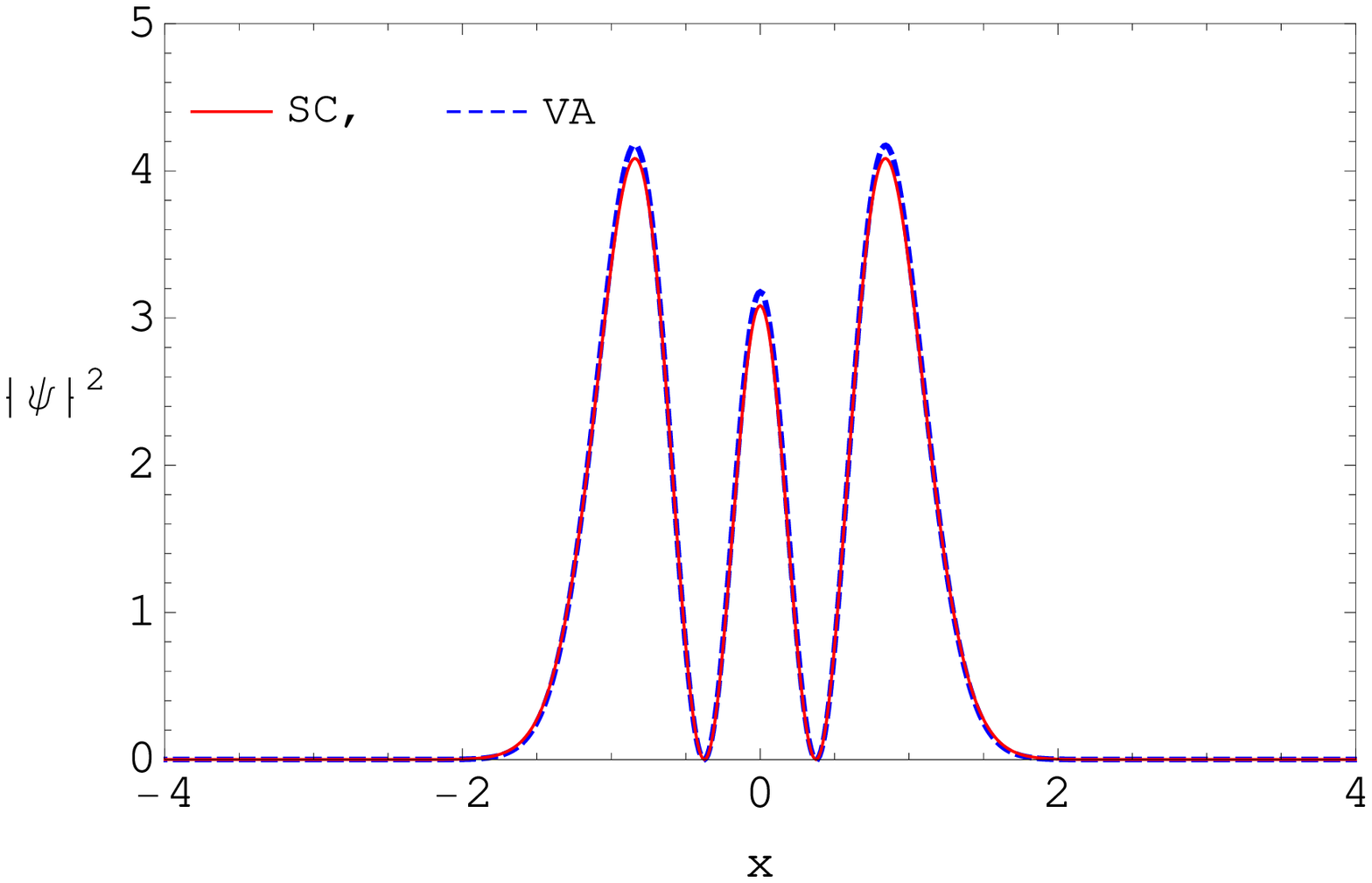}}
\caption{Left panel: The potential $U(a)$ for $g=10$ and $q=0, \,
\pm \, 1$, constructed from VA equations, using the kernel
function (\ref{kernel2}). The inset shows the wave function of a
three-soliton molecule for $q=0$. Right panel: The modulo square
of the wave function, according to VA for parameter values $N=6$,
$\delta=4/\sqrt{\pi}$, $q=0$, $A=1.78$ and $a=0.53$. Dashed line
represents the stationary profile, constructed using the
self-consistent procedure~\cite{salerno2005}.}
\label{fig8}
\end{figure}

First, for analytical convenience we have employed a normalized
Gaussian function for the kernel $R(x)$ in Eq.~(\ref{gpe}) which
does not possess the required long-ranged algebraic decay $\sim
1/x^{3}$ typical of dipolar interactions. On the other hand, an
expression for the dipolar response function for the one-dimensional
setting was derived in \cite{sinha2007}, involving the special
functions. This kernel, however, appears to be quite complicated for
analytical considerations. A more convenient kernel was proposed by
introducing a cutoff parameter $\delta$ \cite{cuevas2009}
\begin{equation}\label{kernel2}
R(x)=\frac{\delta^3}{(x^2+\delta^2)^{3/2}} \ .
\end{equation}
The Eq. (\ref{kernel2}) correctly describes the asymptotic behavior
of dipolar forces, decaying at long distances as $\sim 1/x^3$, and
unlike the response function of Ref. \cite{sinha2007}, does not
feature a cusp at the origin $x=0$. A close similarity between the
two response functions for $\delta = \pi^{-1/2}$ was discussed and
illustrated in \cite{cuevas2009}.

\subsection{Comparison with dipolar model}

We have checked that all the above results are qualitatively
preserved when the calculations are performed with physically more
realistic kernel function~(\ref{kernel2}). The VA results follow
from the same effective Lagrangian  but with the last term in
Eq.~(\ref{lagrangian2}) replaced by $gN\delta^3/(8\pi) \tilde
F(a,\delta)$, where $\tilde F(a,\delta)$ denotes a complicated
function, involving modified Bessel functions and omitted here for
brevity.

In spite of bulky analytical expressions it is possible to solve
the VA equations numerically, and compare the results with the
governing GPE, involving the kernel function~(\ref{kernel2}). In
Fig. \ref{fig8} we show the stationary wave profile of a
three-soliton molecule and the potential curve, obtained using Eq.
(\ref{kernel2}).
\begin{figure}[htb]
\centerline{\includegraphics[width=4.5cm,height=4.5cm,clip]{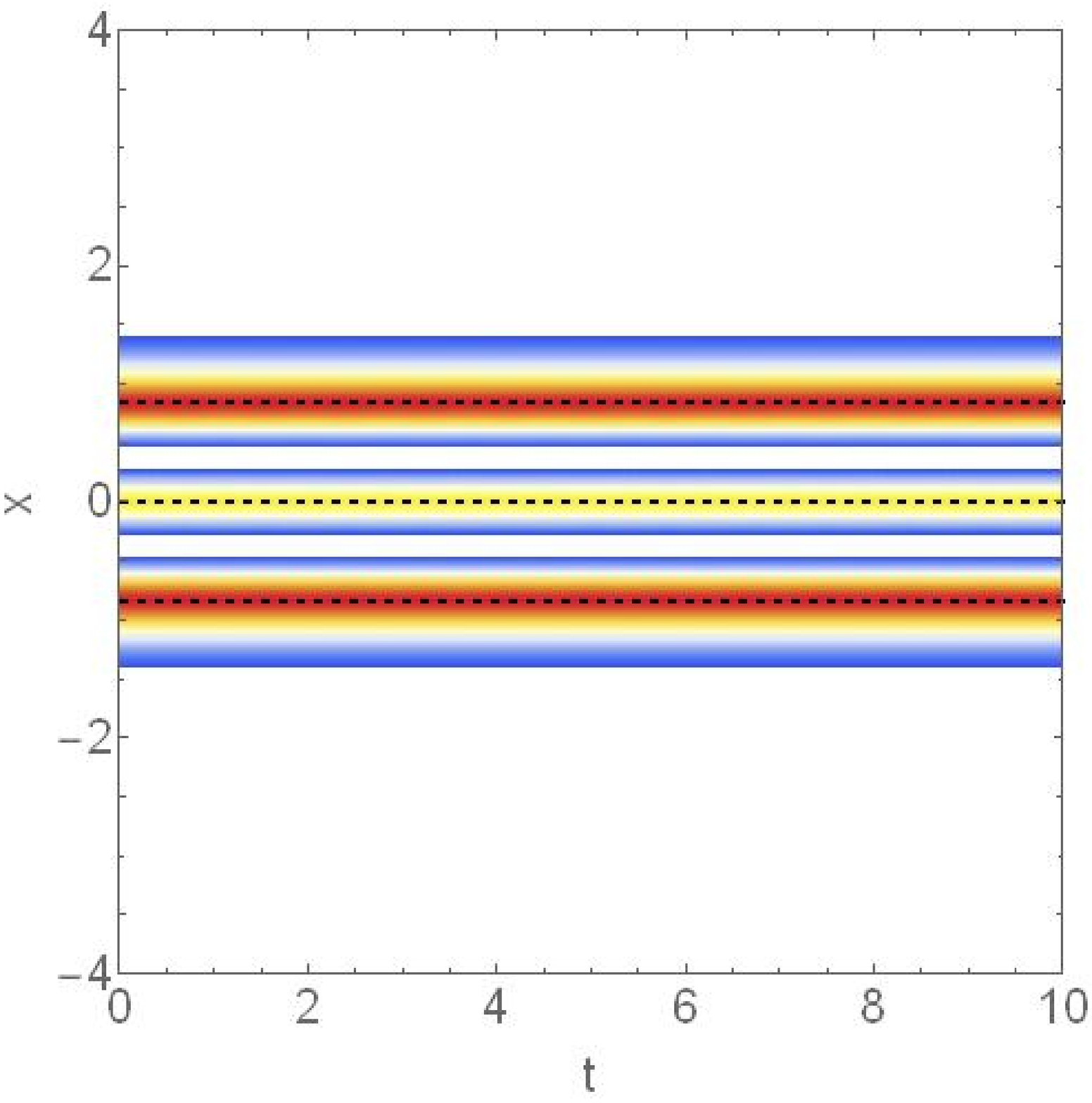}
            \includegraphics[width=4.5cm,height=4.5cm,clip]{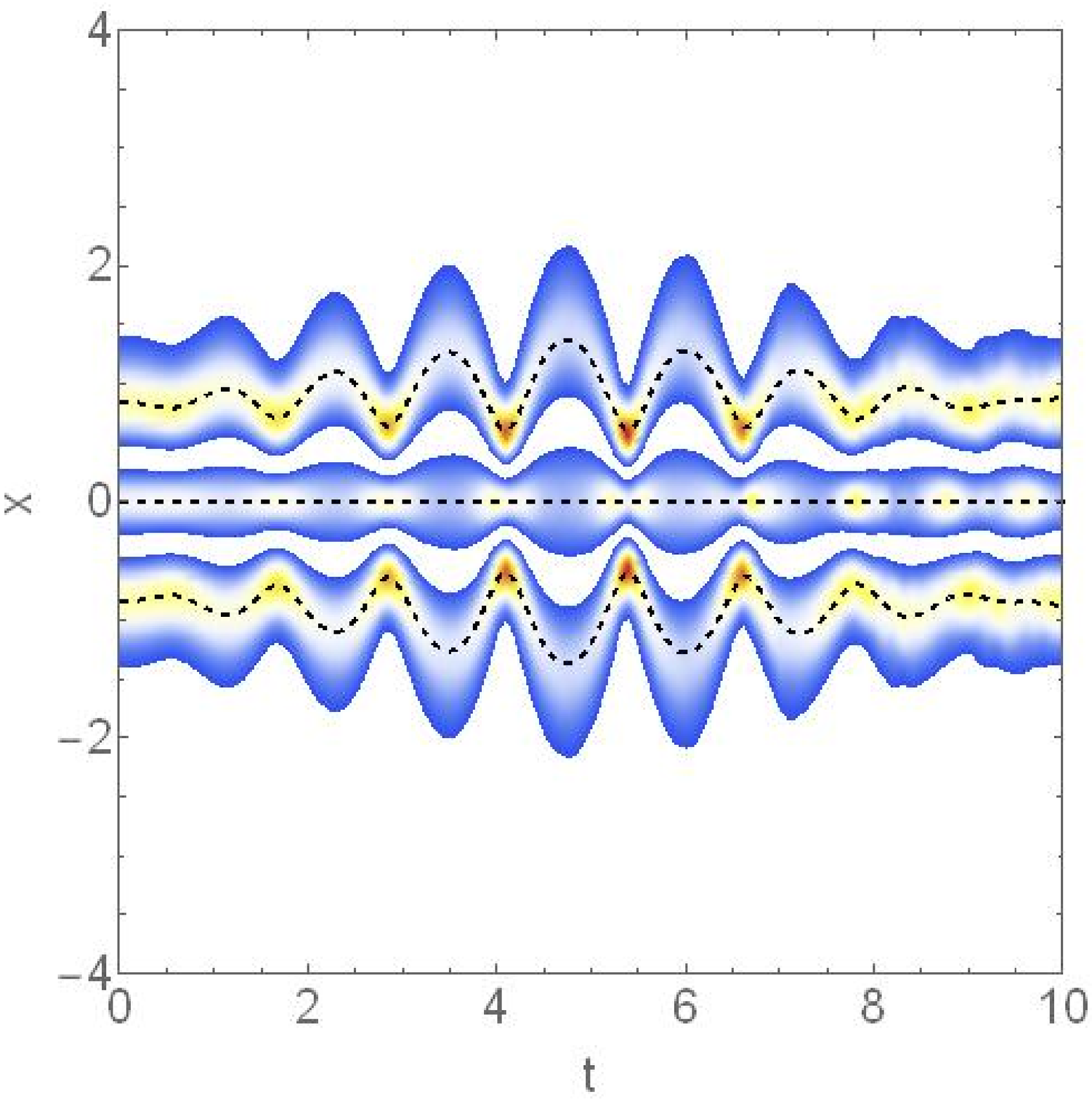}}
\caption{Left panel: Stable propagation of the three-soliton
molecule with parameters predicted by VA for the kernel function
(\ref{kernel2}). The density plot is obtained by numerical
solution of the GPE (\ref{gpe}). Dashed lines correspond to
positions of maxima of the central and lateral solitons $x_m=\pm
\, \sqrt{5/2}\,a(t)$, where the time dependent parameter $a(t)$ is
evaluated from VA. Right panel: Periodic variation of the strength
of dipolar interactions $g(t)~=~g_0(1+\epsilon \sin(\omega_0 t))$
at resonant frequency $\omega_0=5.619$ gives rise to vibration of
lateral solitons with growing amplitude, while the central soliton
remains at origin due to the symmetry. Parameter values: $g_0=10$,
$\epsilon=0.1$. Other parameters are similar to Fig.~\ref{fig8}. }
\label{fig9}
\end{figure}

In Fig. \ref{fig9} we illustrate the GPE and VA dynamics of a
three-soliton molecule under time periodic modulation of the
dipolar interaction. Comparison of Figs. \ref{fig1}-\ref{fig2},
obtained using the Gaussian kernel Eq. (\ref{kernel0}), with Figs.
\ref{fig8}-\ref{fig9}, constructed using the kernel Eq.
(\ref{kernel2}), shows their qualitative similarity. The same is
true for other properties of a soliton molecule discussed in Sec.
III and IV. Thus we conclude that the above phenomena should be
observable in a quasi-1D dipolar BEC.

Second, in our numerical simulations we have neglected by contact
interactions, assuming the dipolar interactions to be dominant.
However, these results survive also in the presence of contact
atomic interactions. The dimensionless quantity, characterizing the
strength of dipole-dipole interactions with respect to contact
interactions, is given~ by
\begin{equation}
\varepsilon = \frac{\mu_0 \mu^2 m}{12 \pi \hbar^2 a_s},
\end{equation}
 where $\mu_0$ is the permeability of vacuum, $\mu, m$ are the
magnetic dipole moment and mass of the atom, respectively, $a_s$ is
the $s$- wave scattering length, responsible for the contact
interactions, and expressed in units of Bohr radius $a_0$. For
electric dipole moments the formula is similar with replacement
$\mu_0 \mu^2 \rightarrow d^2/\epsilon_0$, where $d$ is the electric
dipole moment of atoms, $\epsilon_0$ is the permittivity of vacuum.
For the family of dipolar BECs the estimates are as follows.
$^{52}$Cr: $\mu = 6 \,\mu_B$, $a_s = 16\, a_0$, $\varepsilon_{Cr} =
0.16$. $^{164}$Dy: $\mu = 10 \,\mu_B$, $a_s=92\, a_0$,
$\varepsilon_{Dy} = 1.4$. $^{168}$Er: $\mu = 7\, \mu_B$, $a_s = 60\,
a_0$, $\varepsilon_{Er} = 0.4$. Comparing these values of
$\varepsilon$ with that of the non-dipolar condensate $^{87}$Rb:
$\mu = 1.0\, \mu_B$, $a_s = 0.7\,a_0$, $\varepsilon_{Rb} = 0.007$,
we conclude that the dipolar interactions in Cr, Dy and Er
dominantly contribute to scattering properties of BEC. On the other
hand, when the objective is to observe the dipolar effects clearly,
the contact interactions can be reduced to zero by a magnetic or
optical Feshbach resonance technique~\cite{chen2006}. In the
experiments the setting considered in this paper could be
implemented by applying suitable optical or magnetic fields to
create weakly stretching potentials for the soliton molecule.

In conclusion, we have introduced a three-soliton molecule which
can exist in BEC with nonlocal atomic interactions confined
to quasi-1D traps. The stationary waveform, potential of
inter-soliton interaction, the bond length and some other
characteristic parameters of the three-soliton molecule are
obtained using the variational approach and confirmed by numerical
simulations of the nonlocal GPE. To explore the normal mode
dynamics of the three-soliton molecule, we imposed constant and
non-uniform chirping of the ground state wave function and used
them as initial conditions in numerical simulations of the
nonlocal GPE. We have shown that, contrary to usual triatomic
molecules, the frequency of the asymmetric mode of a three-soliton
molecule is always smaller than the one of the symmetric mode.
Comparison of the frequencies of small amplitude oscillations of
individual solitons, obtained from numerical solution of the
nonlocal GPE, showed a good agreement with the predictions of the
developed model. The results of the present work can be of
interest, e.g. in studies of oscillations of the dipolar BEC over
the surface trap, made of a superconductor
material~\cite{sapina2013,sapina2014}. Normal modes of soliton
molecules can be experimentally measured also in
dispersion-managed optical fibers, where three-soliton molecules
are already produced~\cite{rohrmann2012,rohrmann2013}.

\section*{Acknowledgements}

We thank F. Kh. Abdullaev and E. N. Tsoy for valuable discussions.
M. S. acknowledges partial support from the Ministero
dell'Istruzione, dell'Universit\'a e della Ricerca (MIUR) through
the PRIN (Programmi di Ricerca Scientifica di Rilevante Interesse
Nazionale) grant on ``Statistical Mechanics and Complexity":
PRIN-2015-K7KK8L. B.B.B. thanks the Department of Physics at the
University of Salerno for the hospitality during his visit and
support from the grant $\Phi$A-$\Phi$2-004 of the Agency for
Science and Technologies of Uzbekistan.

\end{document}